\documentclass[a4paper,12pt]{article}
\usepackage{amssymb}
\usepackage{latexsym}

\usepackage{epsfig}
\usepackage{graphicx}

\usepackage{amsmath,amsthm,amsfonts,amssymb}

\newcommand{\be}{\begin{eqnarray}}
\newcommand{\ee}{\end{eqnarray}}
\newcommand{\bea}{\begin{eqnarray}}
\newcommand{\nn}{\nonumber}
\newcommand{\eea}{\end{eqnarray}}

\newcommand{\nk}{\noindent}

\begin{document}

\begin{titlepage}
\begin{flushright}
hep-th/0112019\\ December 2001
\end{flushright}
\begin{centering}
\vspace{.8in}
{\large {\bf Spherically Symmetric Braneworld Solutions with
$^{(4)}R$ term in the Bulk}}
\\

\vspace{.5in} {\bf G. Kofinas\footnote{gkofin@phys.uoa.gr} , E.
Papantonopoulos\footnote{lpapa@central.ntua.gr} , I.
Pappa\footnote{gpappa@central.ntua.gr}}

\vspace{0.3in}
$^{1}$ Department of Physics, Nuclear and Particle Physics
Section, University of Athens,\\Panepistimioupolis GR 157 71,
Ilisia, Athens, Greece
\\
$^{2,\,3}$ Physics Department, National Technical University of Athens,\\
Zografou Campus GR 157 80, Athens, Greece\\
\end{centering}

\vspace{1in}
%%%%%%%%%%%%%%%%%%%%%%% ABSTRACT %%%%%%%%%%%%%%%%%%%%%%%%%%%%%%%%%%%
\begin{abstract}
\nk An analysis of a spherically symmetric braneworld
configuration is performed when the intrinsic curvature scalar is
included in the bulk action; the vanishing of the electric part of
the Weyl tensor is used as the boundary condition for the
embedding of the brane in the bulk. All the solutions outside a
static localized matter distribution are found; some of them are
of the Schwarzschild-$(A)dS_{4}$ form. Two modified
Oppenheimer-Volkoff interior solutions are also found; one is
matched to a Schwarzschild-$(A)dS_{4}$ exterior, while the other
does not. A non-universal gravitational constant arises, depending
on the density of the considered object; however, the conventional
limits of the Newton's constant are recovered. An upper bound of
the order of $TeV$ for the energy string scale is extracted from
the known solar system measurements (experiments). On the
contrary, in usual brane dynamics, this string scale is calculated
to be larger than $TeV$.
\\
\\
\\
\\
\\
\\
\end{abstract}
%%%%%%%%%%%%%%%%%%%%%%%%%%%%%%%%%%%%%%%%%%%%%%%%%%%%%%%%%%%%%%%%%%%%%
\end{titlepage}

\newpage

\baselineskip=18pt
%%%%%%%%%%%%%%%%%% INTRODUCTION %%%%%%%%%%%%%%%%%%%%%%%%%%%%%%%%%%
\section*{1. \,\,\,Introduction}
\hspace{0.8cm} The desire to explore physics beyond the Standard
Model has led us to explore the ideas that spacetime is of a
dimension larger than four, and that we are essentially confined
to a four-dimensional hypersurface. String theories provide a
framework for exploring such ideas, but nevertheless we are still
far away of having a viable low-energy realization of these
theories. Braneworld models consist relevant world realizations in
which some underlying features are often minimized. Replacing, for
example, a whole field with a constant (solitonic solution) may
probably oversimplifies the reality, but at the same time makes it
possible to obtain a more concrete picture, with the hope that any
new behavior appearing will be still present in the more complete
theory. Not only at the cosmological level, but also at a local
one - concerning stars, galaxies, clusters of galaxies - has a
brane solution to be consistent with the various astrophysical
observations, which are often more reliable than the cosmological
ones.
\par
Attempts for obtaining braneworld solutions are cast into two
categories. First, the bulk space assumes a given geometry, a
coordinate system is adopted and the influence on the brane
geometry is somehow extracted. It seems as a disadvantage of this
approach that the bulk is prefixed and also that the brane
imbedding obtained is not gauge-invariant (independent of the
coordinate system chosen). Second, do not specify the exact bulk
geometry, adopt a coordinate system adapted to the brane (gauss
normal coordinates or some relevant one) and deduce a brane
dynamics, containing imprints from the bulk. Assumptions on the
brane geometry are often sufficient for obtaining an exactly
closed brane dynamics. This approach allows of a dynamically
interacting brane with bulk, though this situation is not
necessarily considered. A disadvantage of this method is that
finding a bulk geometry in which the brane consists its boundary
may be a very difficult task. A probable advantage would be the
extraction of common braneworld characteristics holding for a
broad class of bulk backgrounds. In both approaches, if the
codimension is one, Israel matching conditions are necessarily
used. In the present paper we shall elaborate on the second
approach.
\par
The effective brane equations have been obtained \cite{maeda} when
the effective low-energy theory in the bulk is higher-dimensional
gravity. However, a more fundamental description of the physics
that produces the brane could include \cite{sundrum} higher order
terms in a derivative expansion of the effective action, such as a
term for the scalar curvature of the brane, and higher powers of
curvature tensors on the brane. A brane action that contains
powers of the brane curvature tensors has also been used in the
context of the $AdS/CFT$ correspondence (e.g. \cite{maldacena}) to
regularize the action of a bulk $AdS$ space which diverges when
the radius of the $AdS$ space becomes infinite. If the dynamics is
governed not only by the ordinary five-dimensional
Einstein-Hilbert action, but also by the four-dimensional Ricci
scalar term induced on the brane, new phenomena appear. In
\cite{dvali1, dvali2} it was observed that the localized matter
fields on the brane (which couple to bulk gravitons) can generate
via quantum loops a localized four-dimensional worldvolume kinetic
term for gravitons (see also \cite{capper, adler, zee, khuri}).
That is to say, four-dimensional gravity is induced from the bulk
gravity to the brane worldvolume by the matter fields confined to
the brane. It was also shown that an observer on the brane will
see correct Newtonian gravity at distances shorter than a certain
crossover scale, despite the fact that gravity propagates in extra
space which was assumed there to be flat with infinite extent; at
larger distances, the force becomes higher-dimensional. The first
realization of the induced gravity scenario in string theory was
presented in \cite{kiritsis}. Furthermore, new closed string
couplings on Dp-branes for the bosonic string were found in
\cite{corley}. These couplings are quadratic in derivatives and
therefore take the form of induced kinetic terms on the brane. For
the graviton in particular these are the induced Einstein-Hilbert
term as well as terms quadratic in the second fundamental tensor.
Considering the intrinsic curvature scalar in the bulk action, the
effective brane equations have been obtained in \cite{kofinas}.
Results concerning cosmology have been discussed in \cite{collins,
shtanov, nojiri, deffayet, myung}.
\par
The original Randall-Sundrum models \cite{randall}, based on a
Minkowski brane and a specific relation between the bulk
cosmological constant and the brane tension, have drawn much
attention because they might be realizable in supergravity and
superstring compactifications \cite{horava, cvetic, kehagias,
verlinde}. However, any Ricci-flat four-dimensional metric can be
embedded (with the common warped embedding) in $(A)dS_{5}$ (e.g.
\cite{hawking, perry}). This way, a black-string solution
\cite{hawking, emparan, mit, csaki} can be easily constructed.
Furthermore, it is known that any four-dimensional Einstein spaces
can foliate an $(A)dS_{5}$ bulk \cite{hirayama, karch, dewolfe,
kim, kaloper, garriga}. Thus, asymptotically non-flat black holes
(Schwarzschild-$(A)dS_{4}$) can be obtained as slices of the above
precise bulks. Almost all treatments on spherically symmetric
braneworld solutions, as the previously mentioned, representing,
for example, the exterior of a star, do not take care of the
finite extension of the object. Till now, there is no known exact
five-dimensional solution for astrophysical brane black holes.
Furthermore, looking for bulks having some interior star solution
as part of their boundaries is even harder. In \cite{maartens,
wiseman, deruelle}, some interior and exterior solutions were
found, without including the $^{(4)}R$ term.
\par
In the present paper, we discuss the gravitational field of an
uncharged, non-rotating spherically symmetric rigid object when in
the dynamics there is a contribution from the brane intrinsic
curvature invariant. In section 2, we find all the possible
exterior solutions, containing one undetermined parameter which is
the parameter of the Newtonian term. Some of these solutions are
of the Schwarzschild-$(A)dS_{4}$ form. In two cases, we can solve
also the interior problem which reduces to a generalization of the
Oppenheimer-Volkoff solution, and thus determine the unknown
parameter. This is found different from the conventional value of
a localized spherically symmetric distribution within the
framework of four-dimensional general relativity. Hence, a
non-universal Newton's constant, depending on the density of the
object, naturally arises. In section 3, taking care of the
classical experiments of gravity in the solar system, we can set
an upper bound for the five-dimensional Planck mass being of the
order of $TeV$. The revival of the conventional results is
discussed, and also, a comparison with the more standard brane
dynamics is presented. Finally, in section 4 are our conclusions.

%%%%%%%%%%%%%%%%%%%%%% DOMAINWALL MOTION %%%%%%%%%%%%%%%%%%%%%%%%%%%%%%%%%%%%%%%%%%%%%%
\section*{2. \,\,Four-Dimensional Spherically Symmetric Solutions}
\hspace{0.8cm} We consider a 3-dimensional brane $\Sigma$ (with
normal vector field $n^{A}$) embedded in a 5-dimensional spacetime
$M$. Capital Latin letters $A,B,...=0,1,...,4$ will denote full
spacetime, lower Greek $\mu,\nu,...=0,1,...,3$ run over brane
worldvolume, while lower Latin ones span some 3-dimensional
spacelike surfaces foliating the brane, i.e. $i,j,...=1,...,3$.
For convenience, we can quite generally, choose a coordinate $y$
such that the hypersurface $y=0$ coincides with the brane. The
total action for the system is taken to be: \be
S&=&\frac{1}{2\kappa_{5}^{2}}\int_{M}\sqrt{-^{(5)}g}\,\left(^{(5)}R-2\Lambda_{5}\right)d^{5}x+
\frac{1}{2\kappa_{4}^{2}}\int_{\Sigma}\sqrt{-^{(4)}g}\,\left(^{(4)}R-2\Lambda_{4}\right)d^{4}x\nn\\&&+
\int_{M}\sqrt{-^{(5)}g}\,\,L_{5}^{mat}\,d^{5}x+
\int_{\Sigma}\sqrt{-^{(4)}g}\,\,L_{4}^{mat}\,d^{4}x.
\label{action} \ee For clarity, we have separated the cosmological
constants $\Lambda_{5}$, $\Lambda_{4}$ from the rest matter
contents $L_{5}^{mat}$, $L_{4}^{mat}$ of the bulk and the brane
respectively. $\Lambda_{4}/\kappa_{4}^{2}$ can be interpreted as
the brane tension of the standard Dirac-Nambu-Goto action, or as
the sum of a brane worldvolume cosmological constant and a brane
tension. We basically concern on the case with no fields in the
bulk, i.e. $^{(5)}T_{AB}=0$.
\par
From the dimensionful constants $\kappa_{5}^{2}$, $\kappa_{4}^{2}$
the Planck masses $M_{5}$, $M_{4}$ are defined as:
 \be
  \kappa_{5}^{2}=8\pi
G_{(5)}=M_{5}^{-3}\,\,\,\,\,,\,\,\,\,\, \kappa_{4}^{2}=8\pi
G_{(4)}=M_{4}^{-2}, \label{planck} \ee with $M_{5}$, $M_{4}$
having dimensions of (length)$^{-1}$. Then, a distance scale
$r_{c}$ is defined as :
 \be
r_{c}\equiv\frac{\kappa_{5}^{2}}{\kappa_{4}^{2}}=\frac{M_{4}^{2}}
{M_{5}^{3}}\,.
 \label{distancescale}
 \ee
Varying (\ref{action}) with respect to the bulk metric $g_{AB}$,
we obtain the equations \be
^{(5)}G_{AB}=-\Lambda_{5}g_{AB}+\kappa_{5}^{2}\left(^{(5)}T_{AB}+\,^{(loc)}T_{AB}\,\delta(y)\right)\,,
\label{varying}
 \ee
 where
\be
^{(loc)}T_{AB}\equiv-\frac{1}{\kappa_{4}^{2}}\,\sqrt{\frac{-^{(4)}g}
{-^{(5)}g}}\,\left(^{(4)}G_{AB}-\kappa_{4}^{2}\,^{(4)}T_{AB}+
\Lambda_{4}h_{AB}\right)
 \label{tlocal}
\ee is the localized energy-momentum tensor of the brane.
$^{(5)}G_{AB}$, $^{(4)}G_{AB}$ denote the Einstein tensors
constructed from the bulk and the brane metrics respectively.
Clearly, $^{(4)}G_{AB}$ acts as an additional source term for the
brane through $^{(loc)}T_{AB}$. The tensor
$h_{AB}=g_{AB}-n_{A}n_{B}$ is the induced metric on the
hypersurfaces $y=$ constant, with $n^{A}$ the normal vector on
these.
\par
The way the $y$-coordinate has been defined, allows us to write,
at least in the neighborhood of the brane, the 5-line element in
the block diagonal form \be
ds_{(5)}^{2}=-N^{2}dt^{2}+g_{ij}dx^{i}dx^{j}+dy^{2}\,,
\label{lineelement}
 \ee
 where $N,g_{ij}$ are generally functions of $t,x^{i},y$.
 The distributional character of the brane matter content makes
necessary for the compatibility of the bulk equations
(\ref{varying}) the following modified (due to
$^{(4)}G^{\mu}_{\nu}$) Israel-Darmois-Lanczos-Sen conditions
\cite{israel, darmois, lanczos, sen}
 \be
[K_{\nu}^{\mu}]=-\kappa_{5}^{2}\left(^{(loc)}T_{\nu}^{\mu}-
\frac{^{(loc)}T}{3}\delta_{\nu}^{\mu}\right)\,, \label{israel} \ee
where the bracket means discontinuity of the extrinsic curvature
$K_{\mu\nu}=\partial_{y}g_{\mu\nu}/2$ across $y=0$. A
$\mathbf{Z}_{2}$ symmetry on reflection around the brane is
considered throughout.
\par
One can derive from equations (\ref{varying}), (\ref{israel}) the
induced brane gravitational dynamics \cite{kofinas}, which
consists of a four-dimensional Einstein gravity, coupled to a
well-defined modified matter content. More explicitly, one gets
\be
^{(4)}G_{\nu}^{\mu}=\kappa_{4}^{2}\,^{(4)}T_{\nu}^{\mu}-\Big(\Lambda_{4}
+\frac{3}{2}\alpha^{2}\Big)\,\delta_{\nu}^{\mu}+
\alpha\Big(L_{\nu}^{\mu}+\frac{L}{2}\,\delta_{\nu}^{\mu}\Big)\,,
\label{einstein} \ee where $\alpha\equiv 2/r_{c}$, while the
quantities $L^{\mu}_{\nu}$ are related to the matter content of
the theory through the equation \be
L_{\lambda}^{\mu}L_{\nu}^{\lambda}-\frac{L^{2}}{4}\,\delta_{\nu}^{\mu}
=\mathcal{T}_{\nu}^{\mu}-\frac{1}{4}(3\alpha^{2}+2\mathcal{T}_{\lambda}^{\lambda})\,\delta_{\nu}^{\mu}\,,
\label{lll} \ee and $L\equiv L^{\mu}_{\mu}$. The quantities
$\mathcal{T}^{\mu}_{\nu}$ are given by the expression \be
\mathcal{T}_{\nu}^{\mu}&=&\Big(\Lambda_{4}-\frac{1}{2}\,\Lambda_{5}\Big)\delta_{\nu}^{\mu}
-\kappa_{4}^{2}\,^{(4)}T_{\nu}^{\mu}+\nn\\&&
+\frac{2}{3}\,\kappa_{5}^{2}\,\Big(\,^{(5)}\overline{T}\,_{\nu}^{\mu}
+\Big(\,^{(5)}\overline{T}\,_{y}^{y}-\frac{^{(5)}\overline{T}}{4}\Big)\,\delta_{\nu}^{\mu}\Big)
-\overline{\textsf{E}}^{\,\mu}_{\,\nu}\,,
 \label{energy}
 \ee
with $^{(5)}\overline{T}=\,^{(5)}\overline{T}\,_{A}^{A}\,,\,
^{(5)}\overline{T}\,_{B}^{A}= g^{AC}\,^{(5)}\overline{T}\,_{CB}$.
Bars over $^{(5)}T^{A}_{B}$ and the electric part\,
$\textsf{E}^{\,^{\mu}}_{\,\nu}=C^{\mu}_{A \nu B}n^{A}n^{B}$ of the
5-dimensional Weyl tensor $C^{A}_{B C D}$ mean that the quantities
are evaluated at $y=0$. $\overline{\textsf{E}}^{\,\mu}_{\,\nu}$
carries the influence of non-local gravitational degrees of
freedom in the bulk onto the brane \cite{maeda} and makes the
brane equations (\ref{einstein}) not to be, in general, closed.
This means that there are bulk degrees of freedom which cannot be
predicted from data available on the brane. One needs to solve the
field equations in the bulk in order to determine
$\textsf{E}^{\,^{\mu}}_{\,\nu}$ on the brane. In the present
paper, for making (\ref{einstein}) closed, we shall set
$\overline{\textsf{E}}^{\,\mu}_{\,\nu}=0$ as a boundary condition
of the propagation equations in the bulk space. This is somehow
simplified from the viewpoint of geometric complexity, but it is
the first step for investigating the characteristics carried by
the brane curvature invariant on the local brane dynamics we are
interested in. Treatments and solutions without this assumption,
in the context of usual brane dynamics, have been given in
\cite{mit, phpapado, roys, dadhich, maartens, wiseman, deruelle,
casadio}. Due the block-diagonal form of the metric
(\ref{lineelement}) the solution of the algebraic system
(\ref{lll}), whenever \be \mathcal{T}^{i}_{j}=\tau \delta^{i}_{j},
\label{spatial} \ee is \be
L_{0}^{0}=\pm\frac{1}{2\textsf{B}}\,\Big((7-4n_{+}n_{-})\mathcal{T}_{0}^{0}-(3-4n_{+}n_{-})\tau+3\alpha^{2}\Big)\,,
\label{solution2} \ee \be L^{i}_{j}=SE^{i}_{j}\,, \label{ll} \ee
\be L^{0}_{i}=L^{i}_{0}=0\,, \label{l0i} \ee where \be
S=\frac{1}{2\textsf{B}}\left|\mathcal{T}_{0}^{0}+3(\tau+\alpha^{2})\right|\,,
\label{sbb} \ee \be
\textsf{B}=\Big(-6(n_{+}-1)(n_{-}-1)\mathcal{T}_{0}^{0}+2\,n_{+}n_{-}(1-n_{+}n_{-})\tau+\,3(3-2n_{+}n_{-})\alpha^{2}\Big)^{\frac{1}{2}}\,,
\label{b} \ee
 while the matrix $E^{i}_{j}$ is either
$diag(+1,+1,+1)$ (with $n_{+}=3, n_{-}=0$) or $diag(+1,+1,-1)$
(with $n_{+}=2, n_{-}=1$).
\par
Inspecting equation (\ref{einstein}), we see that the inclusion of
the term $^{(4)}R$ has brought a convenient decomposition of the
matter terms. First, standard energy-momentum tensor enters
without having made any choice for the brane tension $\Lambda_{4}$
in terms of $M_{4}, M_{5}$ (in \cite{binetruy, maeda} it has to be
$\Lambda_{4}=3\alpha^{2}/2$). Note that if $^{(4)}R$ is not
included in the action, for $\Lambda_{4}=0$, ordinary
energy-momentum terms cannot arise. Furthermore, in that case,
$\Lambda_{4}$ has to be positive in order for $\kappa_{4}^{2}$ to
be positive. Second, the additional matter terms (which rather
appear here as square roots instead of squares of the
four-dimensional energy-momentum tensor) all contain the factor
$\alpha$ of energy string scale. Thus, conventional
four-dimensional General Relativity revives on some region of a
4-spacetime, whenever these extra terms remain suppressed relative
to the conventional ones; the specific value of $\alpha$
determines the region validity of General Relativity.
\par
From now
on, we are interested on static (non-cosmological) local
braneworld solutions arising from the action (\ref{action}). More
specifically we consider a spherically symmetric line element \be
ds_{(4)}^{2}=-B(r)dt^{2}+A(r)dr^{2}+r^{2}(d \theta^{2}+\sin ^{2}
\theta d \phi^{2}). \label{spherical}
 \ee
The matter content of the 3-universe is a localized spherically
symmetric untilted perfect fluid (e.g. a star) $^{(4)}T_{\mu
\nu}=(\rho+p)u_{\mu}u_{\nu}+pg_{\mu \nu}$ with $\rho=p=0$ for
$r>R$, plus the cosmological constant $\Lambda_{4}$. The matter
content of the bulk is a cosmological constant $\Lambda_{5}$.
These matter contents enter $\mathcal{T}^{\mu}_{\nu}$ in equation
(\ref{energy}) and thus determine $L^{\mu}_{\nu}$ on the right
hand side of our dynamical equations (\ref{einstein}). The result
is
\be
L^{0}_{0}=\pm\frac{1}{2\textsf{B}}\Big(|4\Lambda_{4}-2\Lambda_{5}+
3\alpha^{2}|+\kappa_{4}^{2}\Big((7-3n_{+}n_{-})\rho+(n_{+}+3n_{-})p\Big)\Big)\,,
\label{l00} \ee \be
S=\frac{1}{2\textsf{B}}\Big|4\Lambda_{4}-2\Lambda_{5}+
3\alpha^{2}+\kappa_{4}^{2}\,(\rho-3p)\Big|\,, \label{s00} \ee \be
\textsf{B}=\Big((3-4n_{-})(4\Lambda_{4}-2\Lambda_{5}+
3\alpha^{2})-4\kappa_{4}^{2}\Big(3(n_{-}-1)\rho+(n_{+}-3)p\Big)\Big)^{1/2}\,,
\label{bi} \ee with the only restriction imposed by the square
root appeared in $\textsf{B}$. Thus, necessarily
$4\Lambda_{4}-2\Lambda_{5}+3\alpha^{2}$ is non-negative
(non-positive) for $E_{j}^{i}=\delta_{j}^{i}$ (similarly the other
choice of $E_{j}^{i}$).
\par
For the metric (\ref{spherical}), one evaluates the Ricci tensor
$^{(4)}R_{\mu\nu}$ and then constructs the field equations
(\ref{einstein}). The combination
$^{(4)}R_{rr}/2A\,+\,^{(4)}R_{\theta\theta}/r^{2}\,+\,^{(4)}R_{00}/2B$
provides the following differential equation for $A(r)$ : \be
\left(\frac{r}{A}\right)^{^{\prime}}=1-\kappa_{4}^{2}\,\rho(r)
r^{2}-\Big(\Lambda_{4}+\frac{3}{2}\alpha^{2}\Big)r^{2}+
\frac{\alpha}{2}\Big(3L_{0}^{0}+(n_{+}-n_{-})S\Big)r^{2}\,,
\label{differential} \ee $(^{\prime} \equiv \frac{d}{dr})$.
Eliminating $\frac{A\,^{\prime}}{A}$ from (\ref{differential}) in
the $(\theta\theta)$ component of (\ref{einstein}), we get an
equation for $\frac{B\,^{\prime}}{B}$, from which we obtain \be
\frac{(AB)\,^{^{\prime}}}{AB}=Ar\Big(\kappa_{4}^{2}\,(\rho+p)-
\alpha \Big(L_{0}^{0}+(2-n_{+}+n_{-})S\Big)\Big)\,. \label{AB} \ee
\par
There are various different cases (namely eight) according to the
choice of $E_{j}^{i}$ and the alternative signs in $L_{0}^{0}, S$.
However, in the outside region, there are only four different
cases, according to $n_{+}, n_{-}$ and the $\pm$ sign in
(\ref{l00}); in all these cases, we can integrate equation
(\ref{differential}) in the outside region, obtain the solution
$A_{>}(r)$ and from (\ref{AB}) get the solution $B_{>}(r)$. The
result is \be \frac{1}{A_{>}(r)}=1-\frac{\gamma}{r}-\beta
r^{2}\,\,\,,\,\,\,r\geq R \,, \label{generalA>} \ee \be B_{>}(r)=
\frac{1}{A_{>}(r)}\,F_{n_{+}\,,\,n_{-}}(r)\,\,\,,\,\,\,r\geq R\,,
\label{generalB>} \ee with \be
\beta=\frac{1}{3}\Lambda_{4}+\frac{1}{2}\alpha^{2}-
\alpha\frac{n_{+}-n_{-}\pm
3}{12\sqrt{|3-4n_{-}|}}\sqrt{|4\Lambda_{4}-2\Lambda_{5}+3\alpha^{2}|}\,,
\label{generalbeta} \ee \be
F_{n_{+}\,,\,n_{-}}(r)=1+\Big(f(r)^{\,\alpha r_{1}
\frac{n_{+}-(2+3\sqrt{3})n_{-}}{6\sqrt{3}\,\beta\,(3\gamma-2r_{1})}\sqrt{|4\Lambda_{4}-2\Lambda_{5}+
3\alpha^{2}|}}\,-1\Big)\,\delta_{n_{+}\mp 1\,,\,4-3n_{-}}\,,
\label{F} \ee \be
f(r)=(r-r_{1})\,\Big(\frac{r}{A_{>}}\Big)^{1-\frac{2\gamma}{r_{1}}}\,\,g(r)^{\sqrt{\frac{|r_{1}-\gamma|}{r_{1}+3\gamma}}}\,,
\label{f(r)} \ee where $r_{1}$ is the minimum horizon distance and
$g(r)$ is equal to $\Big|\frac{r\,+\,r_{1}/2\,+\,\sqrt{(r_{1}+3
\gamma)/(4 \beta r_{1})}} {r\,+\,r_{1}/2\,-\,\sqrt{(r_{1}+3
\gamma)/(4\beta r_{1})}}\Big|$ for $\beta
>0$, or $e^{2\arctan \,\,(\sqrt{4|\beta|r_{1}/(r_{1}+3\gamma)}\,\,(r\,+\,r_{1}/2)\,\,)}$
for $\beta<0$. For \,$\beta=0$,
$g(r)^{\sqrt{|r_{1}-\gamma|}\,/\,\beta}$ \,is replaced by
$(r-\gamma)^{4\gamma^{5/2}}\,e^{2\sqrt{\gamma}\,(r+3\gamma)\,(r-\gamma)}$\,.
The $\pm, \,\mp$ signs appearing in (\ref{generalbeta}), (\ref{F})
correspond to the $\pm$ sign of (\ref{l00}). A multiplicative
constant of integration for $B_{>}$ has been absorbed to a
redefinition of time and $\gamma$ is a constant of integration.
Note that for $\beta\leq 0$ there is only one horizon $r_{1}\leq
\gamma$, while for $\beta>0$ (and $27\beta \gamma^{2}<4$ to have
well defined horizons) there are two horizons
$\gamma<r_{1}<3\gamma$ and $1/\sqrt{3\beta}<r_{2}<1/\sqrt{\beta}$.
\par
Solutions (\ref{generalA>}), (\ref{generalB>}) are not yet
completely defined unless the parameter $\gamma$ is determined,
i.e. the interior solution is found. In the case
$E_{j}^{i}=\delta_{j}^{i}$, we can find two situations where
equation (\ref{differential}) does not contain $p$ and so, we can
integrate it in the interior region (we give these solutions
below, equation (\ref{A<})).
\par
As it is seen from (\ref{generalB>}) and (\ref{F}), all the
exterior solutions are either of the form where $A_{>}$, $B_{>}$
are inverse to each other, or of the form where the product
$A_{>}B_{>}$ is equal to $f(r)$ to a power appeared in (\ref{F}).
The first class of these solutions is of the
Schwarzschild-$(A)dS_{4}$ form, while the second is not. For zero
$\beta$ (we can interpret $\beta$ as the effective brane
cosmological constant) the first class of these solutions reduces
to Schwarzschild-like, while the second does not.
Non-Schwarzschild-like exterior solutions were also obtained in
\cite{maartens, phpapado, deruelle, casadio}, but this fact was
rendered to the non-vanishing
$\overline{\textsf{E}}^{\,\mu}_{\,\nu}$. Such irregular behavior
also appears here, due to the intrinsic curvature invariant,
without having involved non-local bulk effects onto the brane.
There is one case of our non-Schwarzschild-$(A)dS_{4}$ solutions
with $\beta>0$, $\gamma/r_{1}=2/3$, where at large distances --
given that the second horizon is actually at cosmological
distances -- $A_{>}B_{>}$ is almost one, i.e. the solution
asymptotes the Schwarzschild-$dS_{4}$.
\par
If we take the covariant derivative (denoted by $_{|}$) with
respect to the induced brane metric $h_{AB}=g_{AB}-n_{A}n_{B}$ of
the equations (\ref{israel}), and make use of the Codacci's
equations, and of the bulk equations (\ref{varying}), we arrive at
the equations $^{(4)}T^{A}_{B|A}=-[^{(5)}T_{CD}]n^{C}h^{D}_{B}$.
When the matter content of the bulk space is only a cosmological
constant, then the common conservation law of our world is
obtained. For the static case we are discussing, this law is
equivalent to the equation \be
\frac{B\,^{\prime}}{B}=-\frac{2p\,^{\prime}}{\rho+p}\,.
\label{continuity} \ee Thus, for $r\leq R$ we get the equation for
$p(r)$ : \be
\frac{p\,^{\prime}}{\rho+p}=\frac{1-A}{2r}-\frac{Ar}{2}
\Big(\kappa_{4}^{2}\,p-\Big(\Lambda_{4}+\frac{3}{2}\alpha^{2}\Big)+\frac{\alpha}{2}L_{0}^{0}-
\frac{3}{2}\alpha \Big(\frac{4}{3}-n_{+}+n_{-}\Big)S\Big)\,.
\label{pe} \ee
\par
We assume a uniform distribution $\rho(r)=\rho_{o}=\frac{3M}{4\pi
R^{3} }$ for $r\leq R$. Then, the immediate integration of
(\ref{continuity}) gives: \be
B_{<}(r)=\frac{\left(1-\frac{\gamma}{R}-\beta
R^{2}\right)F_{n_{+}\,,\,n_{-}}(R)}{\Big(1+\frac{4\pi
R^{3}}{3M}p(r)\Big)^{2}}\,\,\,,\,\,\,r\leq R\,, \label{B<} \ee in
which, continuity of $B(r)$ at $r=R$ and the condition $p(R)=0$
have been used. The vanishing of the pressure at the surface,
which is certainly physically reasonable, is a consequence of the
application of the Israel matching conditions at the stellar
surface \cite{synge, santos}. The pressure $p(r)$ in (\ref{B<}) is
found from (\ref{pe}).
\par
Now, we proceed, as we said before, with the two cases where we
can solve the system of equations (\ref{differential}),
(\ref{pe}). Both have $E^{i}_{j}=\delta^{i}_{j}$. The first case
corresponds to the upper sign of the $\pm$ sign in (\ref{l00}),
and the quantity inside the absolute value of (\ref{s00}) in the
interior of the star being positive. The second case corresponds
to the lower $\pm$ sign in (\ref{l00}) and negative quantity in
(\ref{s00}). In these cases, integration of (\ref{differential})
gives \be
\frac{1}{A_{<}(r)}=1-\Big(\beta+\frac{\gamma}{R^{3}}\Big)r^{2}\,\,\,,\,\,\,r
\leq R\,,\label{A<} \ee where the parameters $\gamma$ and $\beta$
(from (\ref{generalbeta})) are given in terms of $M, \alpha,
\Lambda_{5}, \Lambda_{4}$ by : \newline \underline{First solution}
\be \frac{\gamma}{R^{3}}=\frac{\kappa_{4}^{2}M}{4 \pi R^{3}} +
\frac{\alpha}{2\sqrt{3}}\sqrt{4\Lambda_{4}-2\Lambda_{5}+3\alpha^{2}}
-
\frac{\alpha}{2\sqrt{3}}\sqrt{4\Lambda_{4}-2\Lambda_{5}+3\alpha^{2}+\frac{3\kappa_{4}^{2}M}{\pi
R^{3}}}\,, \label{gamma1} \ee \be
\beta=\frac{1}{3}\Lambda_{4}+\frac{1}{2}\alpha^{2}-
\frac{\alpha}{2\sqrt{3}}\sqrt{4\Lambda_{4}-2\Lambda_{5}+3\alpha^{2}}
\label{beta1} \ee \newline \underline{Second solution} \be
\frac{\gamma}{R^{3}}=\frac{\kappa_{4}^{2}M}{4 \pi R^{3}}+
\frac{\alpha}{2\sqrt{3}}\sqrt{4\Lambda_{4}-2\Lambda_{5}+3\alpha^{2}+\frac{3\kappa_{4}^{2}M}{\pi
R^{3}}}\,, \label{gamma2} \ee \be
\beta=\frac{1}{3}\Lambda_{4}+\frac{1}{2}\alpha^{2}\,.
\label{beta2} \ee The first solution, as it is seen from (\ref{F})
and (\ref{generalB>}), is matched to a Schwarzschild-$(A)dS_{4}$
exterior solution, while the second solution is matched to a
non-Schwarzschild-$(A)dS_{4}$ exterior solution. Note that no
additional constant of integration enters the above solutions by
requiring that the metric is non-degenerate at $r=0$. In the
special case with $4\Lambda_{4}-2\Lambda_{5}+3\alpha^{2}=0$,
(\ref{gamma2}) is matched to an exterior
Schwarzschild-$(A)dS_{4}$.
\par
From (\ref{pe}), $p(r)$ for our two solutions, is found to be \be
p(r)=-\rho_{o}\,\frac{\sqrt{1-
\left(\beta+\frac{\gamma}{R^{3}}\right)r^{2}}\ominus \sqrt{1-
\left(\beta+\frac{\gamma}{R^{3}}\right)R^{2}}}{\sqrt{1-
\left(\beta+\frac{\gamma}{R^{3}}\right)r^{2}}\ominus \omega
\,\sqrt{1- \left(\beta+\frac{\gamma}{R^{3}}\right)R^{2}}}\,,
\label{p<} \ee where \be
\omega^{-1}=1-\frac{2}{\kappa_{4}^{2}\,\rho_{o}}\,\Big(\beta+\frac{\gamma}{R^{3}}\Big)\,\Big(1\mp
\frac{\sqrt{3}\,\alpha}{\sqrt{4\Lambda_{4}-2\Lambda_{5}+3\alpha^{2}+
4\kappa_{4}^{2}\,\rho_{o}}}\Big)^{-1}\,. \label{omega} \ee The
symbol $\ominus$ means $-$, except from the (rather irregular)
case with $\omega<0,\,p>\rho_{o}/|\omega|$, where it becomes $+$.
In the limit $\alpha, \Lambda_{4}\rightarrow 0 $ both solutions
for $A_{<}(r),B_{<}(r),p(r)$ reduce to the known
Oppenheimer-Volkoff solution. Also, in the limit
$\alpha\rightarrow 0$, the exterior solutions corresponding to
(\ref{gamma1}), (\ref{beta1}) and (\ref{gamma2}), (\ref{beta2})
reduce to the Kottler \cite{kottler, gibbons} solution of
4-dimensional General Relativity.
\par
It is of some importance to notice the following. Although three
unrelated parameters $\alpha,\Lambda_{4},\Lambda_{5}$ (which are
rather supposed to be fundamental) enter our problem, the final
exterior solutions contain only two combinations of them, namely
the parameters $\gamma,\beta$. Thus, from exterior experimental
data only two constraints on $\alpha,\Lambda_{4},\Lambda_{5}$ can
be extracted. However, the interior solutions contain,
furthermore, the parameter $\omega$, which means that a third
combination of $\alpha,\Lambda_{4},\Lambda_{5}$ could be obtained
from possible astrophysical information. Thus,
$\alpha,\Lambda_{4},\Lambda_{5}$ can be uniquely determined from
local measurements. Of course, as it is seen from (\ref{omega}),
if the parameters $\alpha, \Lambda_{4}, \Lambda_{5}$ are extremely
small (as will be seen in the next section), the influence of the
bulk effects onto the interior solution is also small.
\par
 It can be seen from
(\ref{gamma1}) that for a given set of parameters $\alpha,
\Lambda_{4}, \Lambda_{5}$, the relative change
$(\gamma/2G_{(4)}M)-1$ on the parameter of the Newtonian term is
negative and it is an increasing function of $\rho_{o}$. This
deviation from the common situation can be interpreted as an
object-dependent gravitational constant, while $M$ remains
unchanged, i.e. $\gamma=2G_{(4)}(\rho_{o})M$, where
$G_{(4)}(\rho_{o})/G_{(4)}=1+2\left(1+\frac{4\Lambda_{4}-2\Lambda_{5}}
{3\,\alpha^{2}}\right)^{-1/2}\,\,\frac{1}{s
\rho_{o}}\left(1-\sqrt{1+s \rho_{o}}\right)$ and $s=32
\pi\,\frac{G_{(4)}}{3
\,\alpha^{2}}\left(1+\frac{4\Lambda_{4}-2\Lambda_{5}}
{3\,\alpha^{2}}\right)^{-1}$. Then, $G_{(4)}(\rho_{o})$ starts
from the value
$G_{(4)}\left(1-\left(1+\frac{4\Lambda_{4}-2\Lambda_{5}}{3\alpha^{2}}\right)^{-1/2}\right)$
when $\rho_{o}\rightarrow 0$, and asymptotically tends to
$G_{(4)}$ for $\rho_{o}\rightarrow \infty$. In this picture,
$G_{N}$, the measured Newton's constant, is not a universal
quantity, but simply corresponds to
$G_{(4)}(\rho_{o,\,everyday})$, where $\rho_{o,\,everyday}$ is the
density of common matter $\sim gr/cm^{3}$. There is a
characteristic value of energy, which can be advocated to these
densities, namely $\alpha_{e}=\sqrt{G_{N}\rho_{o,\,everyday}} \sim
10^{-14}cm^{-1}$. If $4\Lambda_{4}-2\Lambda_{5} \gg 3\alpha^{2}$
(plot 1a in Fig. 1), $G_{(4)}(\rho_{o})$ is always almost equal to
$G_{N}\simeq G_{(4)}$, and no significant deviations from Newton's
constant universality exist. Otherwise (plot 1b in Fig. 1),
significant deviations from $G_{N}$ can arise. Then, there exist
only two situations which do not contradict with the everyday
experience of no deviation from Newton's constant universality.
These are: $\alpha \ll \alpha_{e}$ or $\alpha \gg \alpha_{e}$. In
the first case, $G_{N}\simeq G_{(4)}$ and significant deviations
from $G_{N}$ appear at extremely low densities $\rho_{o}\ll
\alpha^{2}/G_{N}$. In the second case, $G_{N} \simeq
G_{(4)}\left(1-\left(1+\frac{4\Lambda_{4}-2\Lambda_{5}}{3\alpha^{2}}\right)^{-1/2}\right)$
and significant deviations appear at extremely dense objects. In
the next section, we will set upper bounds on $\alpha$, similar to
$\alpha\ll \alpha_{e}$, from solar system experiments, and thus
the second case is excluded. If this is really the situation, the
possibility for the parameters having
$4\Lambda_{4}-2\Lambda_{5}<0$, which leads to repulsive gravity on
very low density objects, is possible. Similar behavior to the
above described, but with extra attraction, appears in the
solution (\ref{gamma2}) (plot 2 in Fig. 1).
%\underline{DIAGRAMS}
%
%
\begin{figure}[h!]
\centering
%\hspace{0.1cm}%
\includegraphics*[width=250pt, height=200pt]{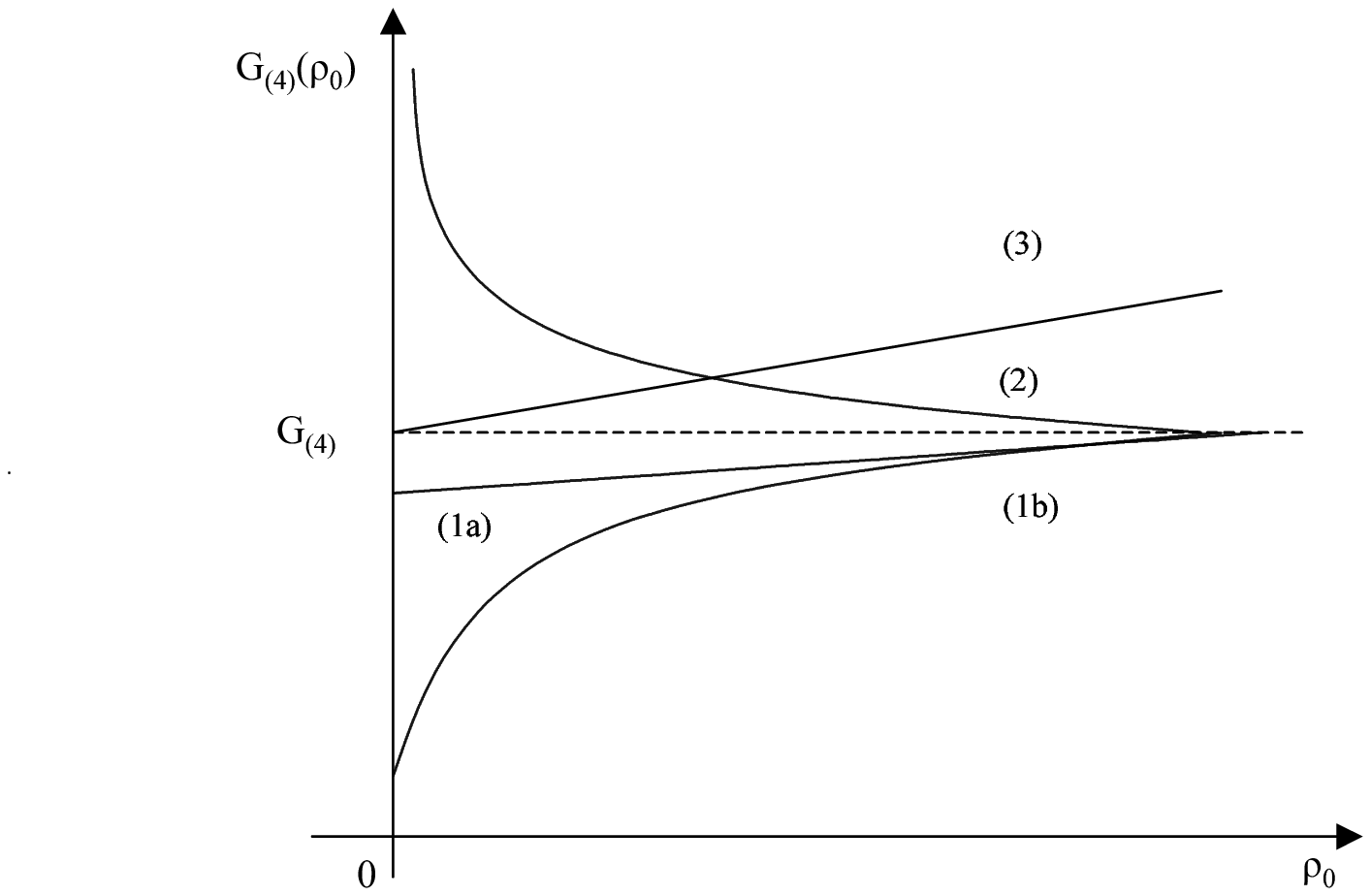}%
\caption{The $\rho_{0}$ dependence of Newton's constant in various
models.}
\end{figure}
\par
Solution (\ref{gamma1}), (\ref{beta1}), being matched to an
exterior Schwarzschild-$(A)dS_{4}$, will be used in the next
section in order to bound the parameters encountered, from
experimental data of our solar system (deflection of light coming
from distant stars, precession of perihelia and radar echo delay).
Since there are two parameters $\beta, \gamma$ in the exterior
solution, connected to the three $\alpha, \Lambda_{4},
\Lambda_{5}$, it is rather necessary to drop one of these three by
hand, in order to get an estimation of the other two. It is
obvious that $\alpha$ cannot be this one, since this is too
restrictive and actually, analyses of this case have been
performed \cite{wright, neupane}. Also, we do not set
$\Lambda_{5}=0$, since then, $\beta$ cannot be negative ($\beta$
negative implies $\Lambda_{5}<0$). Although $\beta$ is the same
quantity which in cosmology plays the role of the effective
cosmological constant \cite{deffayet, kofinas} and it is rather
positive, in the present work, we do not claim any connection with
cosmology, so we would prefer to be able to also deal with a
negative $\beta$. As will be discussed in the next section, this
may be of importance to galactic scale phenomena. In the
following, we choose $\Lambda_{4}=0$. Then, from (\ref{gamma1}),
(\ref{beta1}) we find that for $G_{(4)}\simeq G_{N}$ it is \be
\alpha^{2}=\frac{1}{\gamma
R^{3}}(2G_{N}M-\gamma)(2G_{N}M-\gamma-2\beta R^{3})\,,
\label{alpha} \ee \be \Lambda_{5}=6\beta\,\Big(1-\frac{\beta\gamma
R^{3}}{(2G_{N}M-\gamma)(2G_{N}M-\gamma-2\beta R^{3})}\Big)\,,
\label{Lamda5} \ee while in the other limiting case \be
\alpha^{2}&=&2\beta+\frac{2\gamma^{2}}{R^{3}}
\Big(\gamma-2G_{N}M-\frac{2\gamma}{\beta R^{3}}G_{N}M
\nn\\&&\,\,\,\,\,\,\,\,\,\,\,\,\,\,\,\,\,\,\,\,\,\,\,\,\,\,\,\pm
\sqrt{(2G_{N}M-\gamma)(2G_{N}M-\gamma+\frac{4\gamma}{\beta
R^{3}}G_{N}M)}\,\, \Big)^{-1}\,. \label{alphasmall} \ee
\par
Finally, we note that the non-Schwarzschild-$(A)dS_{4}$ solution
(\ref{gamma2}), (\ref{beta2}) could be also used for extracting
phenomenological bounds on the string parameters from the
solar-system experiments. We have chosen in this paper the
simplest solution for this purpose. However, it is known
\cite{ederys} that the agreement with the solar-system tests of
some metric-based relativistic theory requires on kinematical
grounds that $AB \simeq 1$ to high accuracy in the vicinity of the
sun.

%%%%%%%%%%%%%%%%%%%%%% THE SOLUTION %%%%%%%%%%%%%%%%%%%%%%%%%%%%%%%%%%%%%%%%%%%%%%
\section*{3. \,\,\,Constraints from Classical Tests}
\hspace{0.8cm} A difficulty arising with the calculations of the
measurable quantities (integrals) comes from the fact that the
solution (\ref{gamma1}), (\ref{beta1}) is not asymptotically flat,
but diverges at large distances, thus, an expansion in powers of
$1/r$, performed in the standard PPN (parametrized post-Newtonian)
analysis, does not work here. Hence, one has to make expansions
according to parameters of the problem which are sufficiently
small, and fortunately, such parameters exist.
\par
The motion of a freely falling material particle or photon in a
static isotropic gravitational field (\ref{spherical}) is
described \cite{weinberg} by the equation \be
\Big(\frac{d\phi}{dr}\Big)^{2}=\frac{A_{>}}{r^{4}}\Big(\frac{1}{J^{2}B_{>}}-
\frac{1}{r^{2}}-\frac{E}{J^{2}}\Big)^{-1}\,, \label{path} \ee
where $J,E$ are constants of integration ($E>0$ for material
particles and $E=0$ for photons). At the points of minimum or
maximum distance $r_{o}$, it is $dr/d\phi=0$ and thus \be
J=r_{o}\Big(\frac{1}{B_{>}(r_{o})}-E\Big)^{1/2}\,. \label{J} \ee
\par
We will analyze the three classical solar scale experiments -
deflection of electromagnetic waves coming from distant stars by
the sun, precession of the perihelia of planets, and time delay of
radio waves.
\par
1) \underline{Deflection of light}. Although the metric is not
asymptotically flat, the photon, as it can be seen from equation
(\ref{path}), has $d\phi/dr\rightarrow 0$ as $r\rightarrow
\infty$, and thus, it moves in a ``straight'' line of the
background geometry in that region. The deviation from this line
is measured by the total deflection angle $\Delta \phi
_{d}=2|\phi(r_{o})-\phi(\infty)|-\pi$, where $r_{o}$ is the
minimum distance of the orbit to the sun (when a ray grazes sun
$r_{o}=R_{\odot}$). As it is expected and will be shown below,
$|\beta|$ has an extremely small value, thus, for $\beta>0$ the
horizon $r_{2}$ is of cosmological scale and scattering of light
can be practically defined even in this case. The angular momentum
$J$ is related to the ``impact parameter'' $\textsf{b}$ through
the relation $J=\textsf{b}(1-\beta \textsf{b}^{2})^{-1/2}$.
Integrating (\ref{path}) we arrive at \be
\phi(r)-\phi(\infty)=\sqrt{\frac{r_{o}^{3}}{r_{o}-\gamma}}\int_{r}^{\infty}
\Big[r(r-r_{o})\Big(r^{2}+r_{o}r-\frac{\gamma
r_{o}^{2}}{r_{o}-\gamma}\Big)\Big]^{-1/2}dr\,. \label{intdefl} \ee
For the above expression being well-defined, it has to be
$r_{o}\geq \frac{3}{2}\gamma$ (which is always the case for common
stars such as our sun). Note that the parameter $\beta$ has
disappeared in the expression (\ref{intdefl}), i.e. the deflection
phenomenon is the same as if it had been occurred in a
Schwarzschild field of parameter $\gamma$. Expression
(\ref{intdefl}) leads to an elliptic integral. Since $\gamma$ is
almost $2G_{N}M_{\odot}$ and $r_{o}$ is of the order of
$R_{\odot}$, thus, $\frac{\gamma}{r_{o}}$ is of the order of
$10^{-6}$. Hence, we can expand the integrand of (\ref{intdefl})
to first order in this parameter before integration \cite{edery}.
It is convenient, simultaneously, to set $\sin u
=\frac{r_{o}}{r}$, and the result is \be
\Delta\phi_{d}=\frac{2\gamma}{r_{o}}\,. \label{deflection} \ee
\par
The best measurements on the deflection of light from the sun were
obtained using radio-interferometric methods \cite{fomalont} and
found to be (for $r_{o}=R_{\odot}$) $\Delta\phi_{d}=1.761\pm
0.016$ arc sec. Then, from (\ref{deflection}), \be 29.440\times
10^{4}\,cm<\gamma<29.979\times10^{4}\,cm\,, \label{numbergamma}
\ee which is around the conventional value $2G_{N}M_{\odot}=29.539
\times 10^{4}\,cm$.
\par
2) \underline{Precession of perihelia}. Here, there are two values
$r_{+},r_{-}$ of maximum and minimum distance, satisfying equation
(\ref{J}). The two constants of motion $J,E$ are expressed in
terms of $r_{+},r_{-}$ and are plugged in (\ref{path}). The
expression arising is very complicated, but referring to
\cite{wright}, \cite{neupane}, we can write the precession per
orbit $\Delta\phi_{p}=2|\phi(r_{+})-\phi(r_{-})|-\pi$ as \be
\Delta\phi_{p}=\frac{3\pi\gamma}{L}+\frac{6\pi\beta
L^{3}}{\gamma}\,, \label{precession} \ee where
$L^{-1}=(r^{-1}_{+}+r^{-1}_{-})/2$ is the \textit{semilatus
rectum} of the orbit. Both \cite{wright}, \cite{neupane} agree on
the result (\ref{precession}). Actually, they refer to the
Gibbons-Hawking metric, but their methods are immediately applied
in our case. They disagree on the next order terms, which are,
however, negligible compared to the second term of
(\ref{precession}), for stars with small Schwarzschild radius and
for slightly eccentric orbits.
\par
For Mercury, the uncertainty in the quantity
$\Delta\phi_{p}-\frac{6\pi G_{N}M_{\odot}}{L}$ is $\pm
10^{-9}$\,rad/orbit. Then, taking into account the range
(\ref{numbergamma}) of $\gamma$ received from deflection, we
obtain \be
-7.908\times10^{-43}\,cm^{-2}<\beta<2.465\times10^{-43}\,cm^{-2}\,.
\label{numberbeta} \ee
\par
The bounds (\ref{numbergamma}), (\ref{numberbeta}) give from
equation (\ref{alpha}): \be \alpha<4.379\times10^{-16}\,cm^{-1}\,.
\label{numberalpha} \ee Actually, as far as, the upper bound of
$|\beta|$ remains many orders of magnitude smaller than
$\frac{G_{N}M_{\odot}}{R_{\odot}}$, the above result, as can be
seen from (\ref{alpha}), is insensitive to the exact value of
$\beta$. Furthermore, the fact that $\alpha$ has upper, instead of
lower, bound is due to the specific functional form of the
expression (\ref{alpha}) in terms of $\gamma$. This means that the
crossover scale $r_{c}>4.567\times 10^{15}\,cm$, i.e. the lower
bound of $r_{c}$ is a few times the diameter of our planetary
system. Thus, the five-dimensional fundamental Planck scale
$M_{5}$ is less than $0.9\,TeV$. From equation (\ref{alphasmall}),
one can see that for $\beta\rightarrow 0$, $\alpha\rightarrow 0$
and then, from (\ref{numbergamma}), (\ref{numberbeta}), an upper
bound of the order $10^{-22}\,cm^{-1}$ is set for $\alpha$, which
is incompatible with $\alpha \gg \alpha_{e}$. Thus, this case is
not acceptable.
\newline
From (\ref{Lamda5}), an upper bound for $\Lambda_{5}$ can be
obtained
\be
\Lambda_{5}<3.804\times10^{-43}\,cm^{-2}\,.
\label{numberlamda5} \ee
\par
Uncertainties in the measurement of the precession of perihelion
are known to exist, due to the rotation of sun; thus, it is better
to examine the bounds on $\beta$ from the radar echo delay
independently.
\par
3) \underline{Radar echo delay}. The time required for a radar
signal to go from a point $r$ to the closest to the sun point
$r_{o}$ of its orbit is \be t(r,r_{o})=\int_{r_{o
}}^{r}\Big(\frac{A_{>}}{B_{>}}\Big)^{1/2}
\Big(1-\frac{r_{o}^{2}}{r^{2}}\,\frac{B_{>}}{B_{>}(r_{o})}\Big)^{-1/2}dr\,.
\label{time} \ee As in the deflection of light, expanding to first
order in $\frac{\gamma}{R}$ we obtain \be
t(r,R)&=&\frac{1}{\sqrt{|\beta|}}\arctan h
\Big(\sqrt{|\beta|}\,\sqrt{\frac{r^{2}-R^{2}} {1-\beta
R^{2}}}\Big)+\nn\\&& \gamma \Big(\,\ln \frac{\sqrt{1-\beta
R^{2}}\,r+\sqrt{r^{2}-R^{2}}}{R\sqrt{1-\beta
r^{2}}}+\frac{1}{2\sqrt{1-\beta
R^{2}}}\sqrt{\frac{r-R}{r+R}}\,\,\frac{1+\beta r R}{1-\beta
r^{2}}\Big)\,. \label{time1} \ee This expression holds for
$\beta>0$, while for $\beta<0$, $\arctan h
\left(\sqrt{|\beta|}\,\sqrt{\frac{r^{2}-R^{2}} {1-\beta
R^{2}}}\right)$ has to be replaced by $\frac{\pi}{2}-\arctan
\left(\frac{1}{\sqrt{|\beta|}}\,\sqrt{\frac{1-\beta R^{2}}
{r^{2}-R^{2}}}\right)$. Whenever $|\beta| r^{2}\ll 1$, the above
expression takes the form \be t(r,R) \simeq
\sqrt{r^{2}-R^{2}}+\gamma \,\ln
\frac{r+\sqrt{r^{2}-R^{2}}}{R}+\frac{\gamma}{2}\,\sqrt{\frac{r-R}{r+R}}+
\frac{\beta}{3}\,(r^{2}-R^{2})^{3/2}\,. \label{timeapprox} \ee We
will use this expression to get bounds from the solar radar echo
experiments. Notice, however, that (\ref{time1}) may be applicable
to some more general cases.
\par
In \cite{reasenberg}, the time delay on solar system scales was
measured to an accuracy of 0.1 per cent. A ray that leaves the
Earth, grazes the sun, reaches Mars and comes back would have a
time delay of $248 \pm 0.25\,\mu s$ where the $248 \mu s$ is the
exact prediction of the ``Shapiro'' time delay and the uncertainty
$\pm 0.25 \mu s$ can be used to constrain $\beta$. At superior
conjunction, the radius of the sun to the Earth, $r_{e}$, and to
the Mars, $r_{m}$, are much greater than the radius of the sun
$R_{\odot}$, and thus, $\frac{2}{3}\beta (r_{e}^{3}+r_{m}^{3})=\pm
0.25 \,\mu s$. This constrains $\beta$ to the range \be
|\beta|<7.555\times 10^{-37}\,cm^{-2}\,. \label{betaradar} \ee
\par
It is interesting to compare the bounds on the various parameters
of the brane theory with an $^{(4)}R$ term, with the bounds on the
parameters which are resulting from the brane dynamics without the
$^{(4)}R$ term. In \cite{maeda}, the dynamics on the brane is
given, instead of (\ref{einstein}), by the following equation \be
^{(4)}G^{\mu}_{\nu}&=&
\frac{\kappa_{5}^{4}}{6\kappa_{4}^{2}}\Lambda_{4}\,^{(4)}T^{\mu}_{\nu}-
\frac{1}{2}\Big(\Lambda_{5}+\frac{\kappa_{5}^{4}}{6\kappa_{4}^{4}}\Lambda_{4}^{2}\Big)\,\delta^{\mu}_{\nu}
\nn\\&&-\,\frac{\kappa_{5}^{4}}{24}\Big(6\,^{(4)}T^{\mu}_{\rho}\,^{(4)}T^{\rho}_{\nu}-
2\,^{(4)}T\,^{(4)}T^{\mu}_{\nu}-3\,^{(4)}T^{\rho}_{\sigma}\,^{(4)}T^{\sigma}_{\rho}\,\delta^{\mu}_{\nu}
+\,^{(4)}T^{2}\,\delta^{\mu}_{\nu}\Big)-
\overline{\textsf{E}}^{\,\mu}_{\,\nu}\,. \label{mae} \ee For
$\overline{\textsf{E}}^{\,\mu}_{\,\nu}=0$, following the same
steps for solving (\ref{mae}), as before, we arrive to the unique
Schwarzschild-$(A)dS_{4}$ exterior solution
$B_{>}(r)=\frac{1}{A_{>}(r)}$, where $A_{>}(r)$ is given by
(\ref{generalA>}). The parameters of this solution, denoted by the
subscripts SMS, are given by \be \gamma_{SMS}=
\frac{\kappa_{4}^{2}\,\Lambda_{4,SMS}}{6\pi
\alpha_{SMS}^{2}}\Big(1+\frac{3\kappa_{4}^{2}M}{8\pi
\Lambda_{4,SMS} R^{3}}\Big)M\,, \label{gammaSMS} \ee \be
\beta_{SMS}=\frac{1}{6}\Lambda_{5,SMS}+\frac{\Lambda_{4,SMS}^{2}}{9
\alpha_{SMS}^{2}}\,. \label{betaSMS} \ee It is obvious that the
conventional value $2G_{N}M$ of the Newtonian term can dominate
$\gamma_{SMS}$ only if $\Lambda_{4,SMS}=3\alpha_{SMS}^{2}/2$. This
is the same value which revives the common four-dimensional
energy-momentum terms in the general equation (\ref{mae}). This
value is substituted in (\ref{gammaSMS}), (\ref{betaSMS}) and
then, using the bounds (\ref{numbergamma}), (\ref{numberbeta})
from the classical tests, we can set bounds on $\alpha_{SMS}$,
$\Lambda_{5,SMS}$. More specifically, since $\Lambda_{5,SMS}$ is
not contained in (\ref{gammaSMS}), equation (\ref{numbergamma}) is
enough for finding \be
\alpha_{SMS}>2.425\times10^{-13}\,cm^{-1}\,,
\label{numbergammaSMS} \ee which means $M_{5}>7\,TeV$. Then, from
(\ref{betaSMS}) \be
\Lambda_{5,SMS}<-8.818\times10^{-26}\,cm^{-2}\,,
\label{numberlamda5SMS} \ee and only a bulk of negative curvature
is allowed in this approach. The above results are exact, since
now, there are only two unknown parameters $\alpha_{SMS},
\Lambda_{5,SMS}$ to be determined from the two $\gamma_{SMS},
\beta_{SMS}$. It is seen from equation (\ref{gammaSMS}) (plot 3 in
Fig. 1) that the point particle limit of infinite density cannot
be obtained (contrary to the plots 1a, 1b, 2 of Fig. 1), since
then, $G_{(4)}\rightarrow \infty$. Even for different boundary
conditions \cite{maartens} the above limit, sometimes, is not
defined at all.
\par
Finishing, we make the following comment. In our second solution
(\ref{gamma2}), we have obtained $\gamma>2G_{N}M$. Thus, from
(\ref{deflection}), the deflection angle $\Delta\phi_{d}$ is
larger than the corresponding ``Einstein'' deflection
$\frac{4G_{N}M}{r_{o}}$. This situation of increased deflection
(compared to that caused from the luminous matter) has been well
observed in galaxies or clusters of galaxies, and the above
solution might serve as a possible way for providing an
explanation. In Weyl gravity \cite{kazanas, mannheim, edery}, the
above increase is advocated to some parameter like our $\beta$
(with the difference of a linear instead of quadratic term), which
has to be positive in order to account for this (see also
\cite{ederys, beke}). But, then, a $\beta>0$ cannot account for
the additional attractive force needed to explain the galactic
rotation curves. In our solution, instead, there is the additional
freedom for the parameter $\beta$ being negative, which can be
used for the galactic rotation curve fittings. Notice also that
the Gibbons-Hawking solution cannot explain this way the extra
deflection in galaxies. Alternative gravity theories may probably
not have succeeded their best in illuminating the missing mass
problem, but this does not mean that a new gravity modification
must not be tested in the arena of local phenomena; it is certain
that the whole topic deserves a more thorough investigation.

%%%%%%%%%%%%%%%%%%%%%%%%%%% CONCLUSIONS %%%%%%%%%%%%%%%%%%%%%%%%%%%%%%%%%%%%%%%
\section*{4. \,\,\,Conclusions}
\hspace{0.8cm} In the present paper, we have investigated the
influence of the brane curvature invariant included in the bulk
action, on the local spherically symmetric braneworld solutions.
The brane dynamics is made closed by assuming the vanishing of the
electric part of the Weyl tensor as a boundary condition for the
propagation equations in the bulk space. All the exterior
solutions of a compact rigid object have been obtained. Some of
them are of the Schwarzschild-$(A)dS_{4}$ form. Furthermore, two
generalized interior Oppenheimer-Volkoff solutions have been
found, one of which is matched to a Schwarzschild-$(A)dS_{4}$
exterior, while the other does not. A remarkable consequence is
that the bulk space ``sees'' the finite region of the body and
modifies the parameter of the Newtonian term in the outside
region. No contradiction with the everyday Newton's constant
universality leads to bounds on the string scale. The known
classical solar system tests, which were used in the past to check
the validity of General Relativity, are here used to put precise
bounds on the parameters of our model. More specifically, the
crossover scale is found to be beyond our planetary system
diameter, which means that the upper bound for the energy string
scale is of the order of $TeV$. The limit of the idealized
infinite density point particle is obtained, and significant
deviations from the known Newton's constant might occur on
extremely low density matter distributions. In usual brane
dynamics, contrary to our case, the solar tests impose a lower,
instead of upper, bound of the above order on the string scale.
Furthermore, in that case, for obtaining exterior
non-Schwarzschild-$(A)dS_{4}$ solutions, one has to consider
non-local bulk effects.
\par
We have followed a braneworld viewpoint for obtaining braneworld
solutions, ignoring the exact bulk space. We have not provided a
description of the gravitational field in the bulk space, but
confined our interest to effects that can be measured by
brane-observers. However, our formalism assures the existence of a
5-dimensional Einstein space as the bulk space. By making
assumptions for obtaining a closed brane dynamics, there is no
guarantee that the brane is embeddable in a regular bulk. This is
the case for a Friedmann brane \cite{binetruy}, whose symmetries
imply that the bulk is Schwarzschild-$AdS_{5}$ \cite{muko, bow}. A
Schwarzschild brane can be embedded in a ``black string'' bulk
metric, but this has singularities \cite{hawking, gregory}. The
investigation of bulk backgrounds which reduce to Schwarzschild
(or Schwarzschild-$(A)dS_{4}$) black holes is in progress.
\par
It is clear that a density-dependent gravitational constant,
generally violates, at the weak field limit, Newton's third law of
equal action-reaction. This, furthermore, means violation of the
conservation of linear momentum and incapability to define
precisely the potential energy for a system of two masses. Since
the point particle limit does not meet any problem in our model,
and also the Newtonian limit arises in metric-based theories for
point particles moving along geodesics, we think that the
understanding of the motion of the extended body in General
Relativity (or more generally) would shed light to the above
subteties. Beyond these, in our everyday phenomena, where very low
density distributions do not contribute gravitationally, no such
difficulties arise. However, such situations may be relevant to
early stages of the universe, before or during structure
formation.
\par
As a motivation for further speculation, we refer that it would be
quite interesting, even for the formal status of the theory, if
the existence of the $\rho_{o}\rightarrow \infty$ asymptotic
behavior of the solutions found here, keeps valid, whenever
$^{(4)}R$ term is present. Besides this, it is known that in
cosmology the $^{(4)}R$ term revives the desirable early universe
of standard General Relativity. However, to conclude, as
braneworld solutions are continuously investigated, they have to
confront with the accumulated cosmological and astrophysical
observations, if the underlying theories wish to be considered as
viable generalizations of General Relativity.

%%%%%%%%%%%%%%%%%%%%%%%%%%%% BIBLIOGRAPHY %%%%%%%%%%%%%%%%%%%%%%%%%%%%%%%%%%%%%%

%%%%%%%%%%%%%%%%%%%%%%%%%%%%%%%%%%%%%%%%%%%%%%%%%%%%%%%%%%%%%%%%%%%%%%%%%%%%%%


\begin{thebibliography}{99}

\bibitem{maeda} T. Shiromizu, K. Maeda and M. Sasaki, ``The Einstein equations on the 3-Brane World'', Phys. Rev. D62 (2000) 024012, gr-qc/9910076.

\bibitem{sundrum} R. Sundrum, ``Effective field theory for a three-brane
                  universe'', Phys. Rev. D59 (1999) 085009, hep-ph/9805471.

\bibitem{maldacena} O. Aharony, S.S. Gubser, J. Maldacena, H. Ooguri and Y.
                    Oz, ``Large N field theories, string theory and gravity'', Phys. Rept. 323, 183 (2000), hep-th/9905111.

\bibitem{dvali1} G. Dvali, G. Gabadadze and M. Porati, ``4D Gravity on a Brane in 5D Minkowski Space'', Phys. Lett. B485 (2000) 208, hep-th/0005016.

\bibitem{dvali2} G. Dvali and G. Gabadadze, ``Gravity on a Brane in Infinite Volume Extra Space'', Phys. Rev. D63, 065007 (2001), hep-th/0008054.

\bibitem{capper} D.M. Capper, ``On quantum corrections to the graviton propagator'', Nuovo Cim. A25 (1975) 29.

\bibitem{adler} S.L. Adler, ``Order R vacuum action functional in scalar
                free unified theories with spontaneous scale breaking'', Phys. Rev. Lett. 44 (1980) 1567;
                ``A formula for the induced gravitational constant'', Phys. Lett. B95 (1980) 241;
                ``Einstein gravity as a symmetry breaking effect in quantum field theory'', Rev. Mod. Phys. 54 (1982) 729;
                Erratum-ibid. 55 (1983) 837.

\bibitem{zee} A. Zee, ``Calculating Newton's gravitational constant in infrared
              stable Yang-Mills theories'', Phys. Rev. Lett. 48 (1982) 295.

\bibitem{khuri} N. N. Khuri, ``An upper bound for induced gravitation'', Phys. Rev. Lett. 49 (1982) 513;
                ``The sign of the induced gravitational constant'', Phys. Rev. D26
                (1982) 2664.

\bibitem{kiritsis} E. Kiritsis, N. Tetradis and T.N. Tomaras, ``Thick branes and 4D gravity'', JHEP 0108 (2001) 012, hep-th/0106050.

\bibitem{corley} S. Corley, D. Lowe and S. Ramgoolam, ``Einstein-Hilbert action on the brane for the bulk graviton'',
                 JHEP 0107 (2001) 030, hep-th/0106067.

\bibitem{kofinas} G. Kofinas, ``General brane cosmology with $^{(4)}R$ term in $(A)dS_{5}$ or Minkowski
                  bulk'', JHEP 0108 (2001) 034, hep-th/0108013.

\bibitem{collins} H. Collins and B. Holdom, ``Brane cosmologies without Orbifolds'', Phys. Rev. D62 (2000) 105009, hep-ph/0003173.

\bibitem{shtanov} Y. Shtanov, ``On brane-world cosmology'', hep-th/0005193.

\bibitem{nojiri} S. Nojiri and S.D. Odintsov, ``Brane-World Cosmology
                 in higher derivative gravity or warped compactification
                 in the next-to-leading order of
                 $AdS/CFT$ correspondence'', JHEP 0007 (2000) 049,
                 hep-th/0006232.

\bibitem{deffayet} C. Deffayet, ``Cosmology on a brane in Minkowski
                   bulk'', Phys. Lett. B502 (2001) 199, hep-th/0010186.

\bibitem{myung} N.J. Kim, H.W. Lee and Y.S. Myung, ``Role of the brane
                curvature scalar in the brane world cosmology'',
                Phys. Lett. B504 (2001) 323, hep-th/0101091.

\bibitem{randall} L. Randall, R. Sundrum, ``A large mass hierarchy from a
                  small extra dimension'', Phys. Rev. Lett. 83 (1999) 3370,
                  hep-th/9905221; ``An alternative to compactification'',
                  Phys. Rev. Lett. 83 (1999) 4690, hep-th/9906064.

\bibitem{horava} P. Horava and E. Witten, ``Heterotic and type I string dynamics from eleven dimensions'', Nucl. Phys. B460 (1996) 506, hep-th/9510209.

\bibitem{cvetic} M. Cvetic and H. Soleng, ``Supergravity domain walls'', Phys. Rep. 282 (1997) 159, hep-th/9604090.

\bibitem{kehagias} A. Kehagias, ``Exponential and power-law hierarchies from supergravity'',
                   Phys. Lett. B469 (1999) 123, hep-th/9906204.

\bibitem{verlinde} H. Verlinde, ``Holography and compactification'', Nucl. Phys. B580 (2000) 264, hep-th/9906182.

\bibitem{hawking} A. Chamblin, S.W. Hawking and H.S. Reall, ``Brane-world black
                  holes'', Phys. Rev. D61 (2000) 065007, hep-th/9909205.

\bibitem{perry} D. Brecher and M.J. Perry, ``Ricci-Flat Branes'',
                Nucl. Phys. B566 (2000) 151, hep-th/9908018.

\bibitem{emparan} R. Emparan, G.T. Horowitz and R.C. Myers, ``Exact description of black holes on
                  branes'', JHEP 0001 (2000) 007, hep-th/9911043.

\bibitem{mit} A. Chamblin, H.S. Reall, H. Shinkai and T. Shiromizu,
                   ``Charged brane-world black holes'', Phys. Rev. D63 (2001) 064015, hep-th/0008177.

\bibitem{csaki} A. Chamblin, C. Csaki, J. Erlich and T.J.
                 Hollowood, ``Black diamonds at brane junctions'',
                 Phys. Rev. D62 (2000) 044012, hep-th/0002076.

\bibitem{hirayama} T. Hirayama and G. Kang, ``Stable black strings in Anti-de Sitter space'', Phys. Rev. D64
                  (2001) 064010, hep-th/0104213.

\bibitem{karch} A. Karch and L. Randall, ``Locally localized
                gravity'', JHEP 0105 (2001) 008, hep-th/0011156.

\bibitem{dewolfe} O. Dewolfe, D.Z. Freedman, S.S. Gubser and
                  A. Karch, ``Modeling the fifth dimension with
                  scalars and gravity'', Phys. Rev. D62 (2000) 046008,
                  hep-th/9909134.

\bibitem{kim} H.B. Kim and H.D. Kim, ``Inflation and gauge hierarchy
              in Randall-Sundrum compactification'', Phys. Rev. D61 (2000)
              064003, hep-th/9909053.

\bibitem{kaloper} N. Kaloper, ``Bent domain walls as
                  braneworlds'', Phys. Rev. D60 (1999) 123506,
                  hep-th/9905210.


\bibitem{garriga} J. Garriga and M. Sasaki, ``Brane-world creation
                  and black holes'', Phys. Rev. D62 (2000) 043523,
                  hep-th/9912118.

\bibitem{maartens} C. Germani and R. Maartens, ``Stars in the
                   braneworld'', Phys. Rev. D64 (2001) 124010, hep-th/0107011.

\bibitem{wiseman} T. Wiseman, ``Relativistic stars in Randall-Sundrum
                  gravity'', Phys. Rev. D65 (2002) 124007, hep-th/0111057.

\bibitem{deruelle} N. Deruelle, ``Stars on branes: the view from the brane'',
                   gr-qc/0111065.

\bibitem{israel} W. Israel, ``Singular hypersurfaces and thin shells in
                 general relativity'', Nuovo Cimento 44B [Series 10]
                 (1966) 1; Errata-$\textit{ibid}$ 48B [Series 10] (1967) 463.

\bibitem{darmois} G. Darmois, ``M$\acute{e}$morial des sciences
                  math$\acute{e}$matiques XXV'' (1927).

\bibitem{lanczos} K. Lanczos, ``Untersuching $\ddot{u}$ber fl$\ddot{a}$chenhafte verteiliung
                 der materie in der Einsteinschen gravitationstheorie'' (1922),
                 unpublished;
                 ``Fl$\ddot{a}$chenhafte verteiliung
                 der materie in der Einsteinschen gravitationstheorie'', Ann. Phys. (Leipzig) 74 (1924)
                 518.

\bibitem{sen} N. Sen, ``$\ddot{U}$ber dei grenzbedingungen des schwerefeldes an unstetig keitsfl$\ddot{a}$chen'', Ann. Phys. (Leipzig) 73 (1924) 365.

\bibitem{phpapado} N. Dadhich, R. Maartens, P. Papadopoulos and V. Rezania,
                   ``Black holes on the brane'', Phys. Lett. B487 (2000)
                   1, hep-th/0003061.

\bibitem{roys} R. Maartens, ``Cosmological dynamics on the
               brane'', Phys. Rev. D62 (2000) 084023,
               hep-th/0004166; ``Geometry and dynamics of the
               brane-world'', gr-qc/0101059.

\bibitem{dadhich} N. Dadhich, ``Negative energy condition and black holes on the
                  brane'', Phys. Lett. B492 (2000) 357,
                  hep-th/0009178; P. Singh and N. Dadhich, ``Non-conformally flat bulk spacetime and the 3-brane
                  world'', Phys. Lett. B511 (2001) 291, hep-th/0104174.

\bibitem{casadio} R. Casadio, A. Fabbri and L. Mazzacurati, ``New black holes
                  in the brane-world?'', Phys. Rev. D65 (2002) 084040, gr-qc/0111072.

\bibitem{binetruy} P. Bin$\acute{e}$truy, C. Deffayet, U.
                   Ellwanger and D. Langlois, ``Brane cosmological evolution in a bulk
                   with cosmological constant'', Phys. Lett. B477 (2000) 285, hep-th/9910219.

\bibitem{synge} J.L. Synge, ``Relativity: the General Theory'',
                North Holland, Amsterdam (1971).

\bibitem{santos} N.O. Santos, Mon. Not. R. Ast. Soc. 216 (1985) 403.

\bibitem{kottler} F. Kottler, Ann. Phys. (Leipzig) 56 (1918) 410.

\bibitem{gibbons} G.W. Gibbons and S.W. Hawking, ``Cosmological event horizons,
                  thermodynamics, and particle creation'', Phys. Rev. D15 (1977) 2738.

\bibitem{wright} E.L. Wright, ``Interplanetary measures can not bound the
                 cosmological constant'', UCLA-ASTRO-ELW-98-01, astro-ph/9805292.

\bibitem{neupane} I.P. Neupane, ``Planetary perturbation with cosmological
                  constant'', gr-qc/9902039.

\bibitem{ederys} A. Edery, ``The bright side of dark matter'',
                Phys. Rev. Lett. 83 (1999) 3990, gr-qc/9905101.

\bibitem{weinberg} S. Weinberg, ``Gravitation and Cosmology'', John Wiley and Sons Inc., 1972.

\bibitem{edery} A. Edery and M.B. Paranjape, ``Classical
                tests for Weyl gravity: deflection of light and time
                delay'', Phys. Rev. D58 (1998) 024011, astro-ph/9708233.

\bibitem{fomalont} E.B. Fomalont and R.A. Sramek, Phys. Rev. Lett. 36 (1976)
                   1475.

\bibitem{reasenberg} R.D. Reasenberg et all., Ap. J. 234 (1979) L219.

\bibitem{kazanas} P.D. Mannheim and D. Kazanas,
                  ``Exact vacuum solution to conformal Weyl
                  gravity and galactic rotation curves'', Ap. J. 342
                  (1989) 635; ``Solutions to the Kerr and
                  Kerr-Newman problems in fourth order conformal Weyl
                  gravity'', Phys. Rev. D44 (1991) 417.

\bibitem{mannheim} P.D. Mannheim,
                   ``Linear potentials and galactic rotation curves - formalism'',
                   astro-ph/9307004;
                   ``Microlensing, Newton-Einstein gravity and conformal
                    gravity'', astro-ph/9412007;
                    ``Attractive and repulsive gravity'', Found. Phys. 30 (2000)
                    709, gr-qc/0001011.

\bibitem{beke} J.D. Bekenstein, M. Milgrom and R.H. Sanders, ``Comment
               on the ``The bright side of dark matter'''',
               Phys. Rev. Lett. 85 (2000) 1346, astro-ph/9911518.

\bibitem{muko} S. Mukoyama, T. Shiromizu and K. Maeda, ``Global structure
               of exact cosmological solutions in the brane
               world'', Phys. Rev. D62 (2000) 024028, hep-th/9912287.

\bibitem{bow} P. Bowcock, C. Charmousis and R. Gregory,
              ``General brane cosmologies and their global spacetime
              structure'', Class. Quant. Grav. 17 (2000) 4745, hep-th/0007177.

\bibitem{gregory} R. Gregory, ``Black string instabilities in anti-de
                  Sitter space'', Class. Quant. Grav. 17 (2000) L125,
                  hep-th/0004101.




 \end{thebibliography}
 \end{document}